\documentclass[accepted,single,9pt]{gipaper-mod}
\usepackage{graphicx}
\usepackage{times}
\usepackage{ifthen}
\usepackage{fancyhdr}
\usepackage{amsmath} 

\newcommand{\figureTopBot}[1]{
  \begin{figure}[!t]{\sloppy #1}\end{figure}
}

\newcommand{\figureTop}[1]{
  \begin{figure}[!t]{\sloppy #1}\end{figure}
}

\newcommand{\ith}{\mbox{$i$-th }}
\newcommand{\jth}{\mbox{$j$-th }}

\newcommand{\Caption}[1]{
  \caption{
    \setlength{\baselineskip}{0.85 \baselineskip}
    \textit{#1}
  }
}

\newcommand{\Abstract}[1]{
  \abstract{
    \setlength{\baselineskip}{0.85 \baselineskip}
    #1
  }
}

\newcommand{\AnonymousAuthors}{0}

\title{Automatic Joint Parameter Estimation from Magnetic Motion Capture Data}

\ifthenelse{\AnonymousAuthors = 0}{ 
  \newauthor{jo}{James F. O'Brien}{}
  \newauthor{bb}{Robert E. Bodenheimer, Jr.}{}
  \newauthor{gb}{Gabriel J. Brostow}{}
  \newauthor{jh}{Jessica K. Hodgins}{}

  \affiliation{~\\
    College of Computing and Graphics, Visualization, and Usability Center \\
    Georgia Institute of Technology \\
    801 Atlantic Drive\\
    Atlanta, GA 30332-0280\\
    e-mail: {\tt [obrienj$|$bobbyb$|$brostow$|$jkh]@cc.gatech.edu}
  }
}{
  \newauthor{jo}{}{}
  \newauthor{bb}{}{}
  \newauthor{gb}{}{}
  \newauthor{jh}{}{}

  \affiliation{~\\
    ~\\
    ~\\
    ~\\
    ~\\
    ~
  }
}

\Abstract{
  This paper describes a technique for using magnetic motion capture
  data to determine the joint parameters of an articulated hierarchy.
  This technique makes it possible to determine limb lengths, joint
  locations, and sensor placement for a human subject without external
  measurements.  Instead, the joint parameters are inferred with high
  accuracy from the motion data acquired during the capture session.
  The parameters are computed by performing a linear least squares fit
  of a rotary joint model to the input data.  A hierarchical 
  structure for the articulated model can also be determined in
  situations where the topology of the model is not known.  Once the
  system topology and joint parameters have been recovered, the
  resulting model can be used to perform forward and inverse kinematic
  procedures.  We present the results of using the algorithm on human
  motion capture data, as well as validation results obtained with
  data from a simulation and a wooden linkage of known dimensions.
}

\begin{document}
  \setlength{\baselineskip}{0.94 \baselineskip}

\begin{keywords}
  \setlength{\baselineskip}{0.85 \baselineskip}
  Animation, Motion Capture, Kinematics, Parameter Estimation,
  Joint Locations, Articulated Figure, Articulated Hierarchy. 
\end{keywords}


\pagestyle{fancy}
\lhead{}
\rhead{\raisebox{.4in}[0in]{\thepage}}
\chead{}
\lfoot{}
\cfoot{
\raisebox{-.2in}[0in]{\includegraphics[height=1cm]{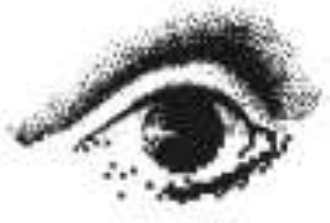}}
~~~~~{\large Graphics \textsf{\textbf{Interface} 2000}~~~Author's Preprint Copy}~~~~~
\raisebox{-.2in}[0in]{\includegraphics[height=1cm]{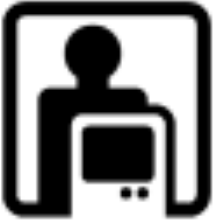}}
}
\rfoot{}
\renewcommand{\headrulewidth}{0pt}
\renewcommand{\footrulewidth}{0pt}


\section{Introduction} 

Motion capture has proven to be an extremely useful technique for
animating human and human-like characters.  Motion capture data
retains many of the subtle elements of a performer's style thereby
making possible digital performances where the subject's unique style
is recognizable in the final product.  Because the basic motion is
specified in real-time by the subject being captured, motion capture
provides a powerful solution for applications where animations with
the characteristic qualities of human motion must be generated
quickly.  Real-time capture techniques can be used to create immersive
virtual environments for training and entertainment applications.

\figureTop{
  \centerline{\includegraphics[width=\columnwidth]{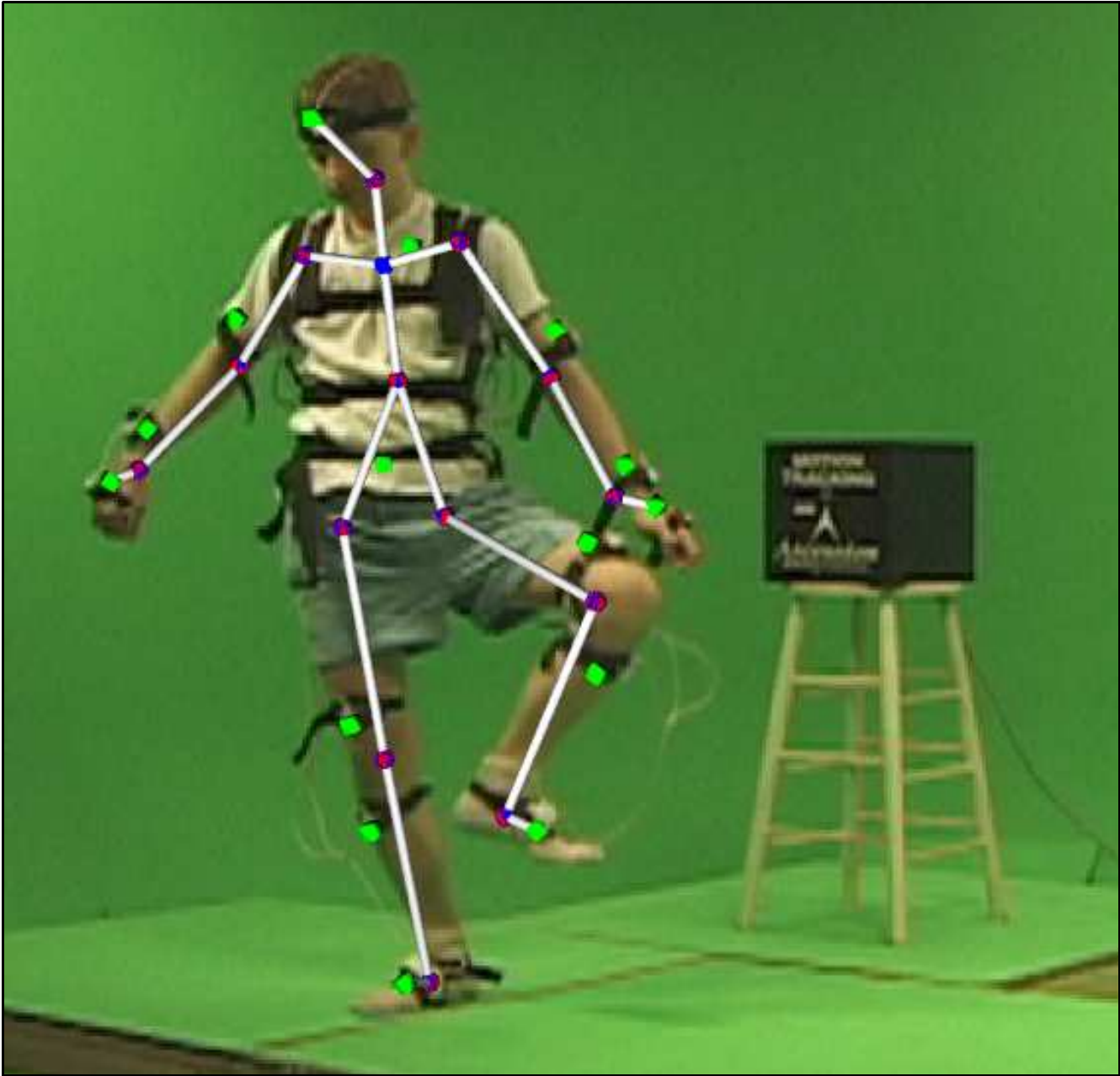}}
  \vskip -0.05in
  \Caption{
     Test subject and generated model. The subject is wearing the
     motion capture equipment during a capture session; the
     superimposed skeletal model is generated automatically from the
     acquired motion capture data. The chest and pelvis sensors are
     located on the subject's back.
  }\label{fig:PersonFigure}
  \vskip -0.2in
}

Although motion capture has many advantages and commercial systems are
improving rapidly, the technology has drawbacks.  Both optical and
magnetic systems suffer from sensor noise and require careful
calibration\cite{Delaney:1998:ott}.  Additionally, measurements such
as limb lengths or the offsets between the sensors and the joints are
often required.  This information is usually gathered by measuring the
subject in a reference pose, but hand measurement is tedious and prone
to error.  It is also impractical for such applications as
location-based entertainment where the delay and physical contact with
a technician would be unacceptable.

The algorithm described in this paper addresses the problem of
calibration by automatically computing the joint locations for an
articulated hierarchy from the global transformation matrices of
individual bodies.  We take motion data acquired with a magnetic
system and determine the locations of the subject's joints and the
relative sensor locations without external measurement.  The technique
imposes no constraints on the sensor positions beyond those necessary
for accurate capture, nor does it require the subject to pose in
particular configurations.  The only requirement is that the data must
exercise all degrees of freedom of the joints if the technique is to
return an unambiguous answer.  Figure~\ref{fig:PersonFigure} shows a
subject wearing magnetic motion capture sensors and the skeletal model
that was generated from the motion data in an automatic fashion.

Intuitively, the algorithm proceeds by examining the sequences of
transformation data generated by pairs of sensors and determining a
pair of points (one in the coordinate system of each sensor) that
remain collocated throughout the sequence.  If the two sensors are
attached to a pair of objects that are connected by a rotary joint, 
then a single point, the center of the joint, fulfills this
criterion.  Errors such as sensor noise and the fact that human joints
are not perfect rotary joints, prevent an exact solution. The       
algorithm solves for a best-fit solution and computes the residual
error that describes how well two bodies ``fit'' together.  This
metric makes it possible to infer the body hierarchy directly from the
motion data by building a minimum spanning tree that treats the residuals
as edge weights between the body parts.

In the following sections, we describe related work in the fields of
graphics, biomechanics, and robotics, and our method for computing the
joint locations from motion data.  We present the results of
processing human motion capture data, as well as validation results
using data from a simulation and from a wooden linkage of known
dimensions.


\section{Background} \label{sec:Background}

Computer graphics researchers have explored various techniques for
improving the motion capture pipeline including parameter estimation
techniques such as the algorithm described in this paper.  Our
technique is closely related to the work of Silaghi and
colleagues\cite{Silaghi:1998:lgs} for identifying an anatomic skeleton
from optical motion capture data.  With their method, the location of
the joint between two attached bodies is determined by first
transforming the markers on the outboard body to the inboard coordinate
system.  Then, for each sensor, a point that maintains an approximately
constant distance from the sensor throughout the motion sequence is
found.  The joint location is determined from a weighted average of
these points.  The sensor weights are determined manually, and because
the coordinate system for the inboard body is not known it must be
estimated from the optical data.  Our technique takes advantage of the
orientation information provided by magnetic sensors.  The computation
is symmetric with respect to the joint between two bodies and does not
require any manual processing of the data.

Inverse kinematics are often used to extract joint angles from global
position data. In the animation community, for example, Bodenheimer
and colleagues\cite{Bodenheimer:1997:tpo} discussed how to apply
inverse kinematics in the processing of large amounts of motion
capture data using a modification of a technique developed by Zhao and
Badler\cite{Zhao:1994:ikp}.  The method presented here is not
an inverse kinematics technique: inverse kinematics assumes that the
dimensions of the skeleton for which joint angles are being computed
is known. Our work extracts those dimensions from the motion capture
data, and thus could be viewed as a preliminary step to an inverse
kinematics computation.

Outside of graphics, the problem of determining a system's kinematic 
parameters from the motion of the system has been studied by researchers 
in the fields of biomechanics\cite{Panjabi:1982:ekp,Panjabi:1982:eca} and
robotics\cite{Karan:1994:cam}.  Biomechanicists are interested in this
problem because the joints play a critical role in understanding the
mechanics of the human body and the dynamics of human motion.
However, human joints are not ideal rotary joints and therefore do not
have a fixed center of rotation.  Even joints like the hip which are
relatively close approximations to mechanical ball and socket joints
exhibit laxity and variations due to joint loading that cause changes
in the center of rotation during movement.  Instead, the parameter
that is often measured in biomechanics is the instantaneous center of
rotation, which is defined as the point of zero velocity during
infinitesimally small motions of a rigid body.

To compute the instantaneous center of rotation, biomechanicists put
markers on each limb and use measurements from various configurations
of the limbs.  To reduce error, multiple markers are placed on each
joint and a least squares fit is used to filter the redundant marker
data\cite{Challis:1995:pdr}.  Spiegelman and Woo proposed a method for
planar motions\cite{Spiegelman:1987:rbm}, and this method was extended to
general motion by Veldpaus and
colleagues\cite{Veldpaus:1988:lsa}. Their algorithm uses multiple
markers on a body measured at two instants in time to establish the
center of rotation.

We are primarily concerned with creating animation rather than
scientific studies of human motion, and our goals therefore differ
from those of researchers in the biomechanics community.  In
particular, because we will use the recorded motion to drive an
articulated skeleton that employs only simple rotary joints, we need
joint centers that are a reasonable approximation over the entire
sequence of motion as opposed to an instantaneous joint center that is
more accurate but describes only a single instant of motion.

The biomechanics literature also provides insight into the errors
inherent in a joint estimation system and suggests an upper bound on
the accuracy that we can expect.  Because the joints of the human body
are not perfect rotary joints, the articulated models used in
animation are inherently an approximation of human kinematics.  Using
five male subjects with pins inserted in their tibia and femur,
Lafortune and colleagues found that during a normal walk cycle the
joint center of the knee compressed and pulled apart by an average of
7\,mm, moved front-to-back by 14.3\,mm, and side-to-side by
5.6\,mm\cite{Lafortune:1992:tdk}.  Another source of error arises
because we cannot attach the markers directly to the bone.  Instead,
they are attached to the skin or clothing of the subject.  Ronsky and
Nigg reported up to 3\,cm of skin movement over the tibia during
ground contact in running\cite{Nigg:1998:bms}.

Roboticists are also interested in similar questions because they need
to calibrate physical devices.  A robot may be built to precise
specifications, but the nominal parameters will differ from those of
the actual unit.  Furthermore, because a robot is made of physical
materials that are subject to deformation, additional degrees of
freedom may exist in the actual unit that were not part of the design
specification.  Both of these differences can have a significant
effect on the accuracy of the unit and compensation often requires
that they be measured\cite{Karan:1994:cam}.  Taking these measurements
directly can be extremely difficult so researchers have developed
various automatic calibration techniques.

The calibration techniques relevant to our research infer these
parameters indirectly by measuring the motion of the robot.  Some of
these techniques require that the robot perform specific actions such
as exercising each joint in
isolation\cite{Zhuang:1993:lsk,Liscano:1985:ikm} or that it assume a
particular set of configurations\cite{Kim:1991:icr,Borm:1989:eso}, and
are therefore not easily adapted for use with human performers.  Other
methods allow calibration from an arbitrary set of configurations but
focus explicitly on the relationship between the control parameters
and the end-effector.  Although our technique fits into the general
framework described by Karan and Vukobratovi\'{c} for estimating
linear kinematic parameters from arbitrary
motion\cite{Karan:1994:cam}, the techniques are not identical because
we are interested in information about the entire body rather than
only the end-effectors.  In addition, we can take advantage of the
position and orientation information provided by the magnetic motion
sensors whereas robotic calibration methods are generally limited to
the information provided by joint sensors (that may themselves be part
of the set of parameters being calibrated) and position markers on the
end-effector.


\vspace*{-0.1in}
\section{Methods}

For a system of $m$ rigid bodies, let~${\cal T}^{i \rightarrow j}$ be
the transformation from the \ith body's coordinate system to the
coordinate system of the \jth body ($i,j \in [0..m-1]$).  The
index $\omega \not\in [0..m-1]$ is used to indicate the
world coordinate system so that~${\cal T}^{i \rightarrow \omega}$ is
the global transformation from the \ith body's coordinate system to
the world coordinate system.

\pagebreak
A transformation ${\cal T}^{i \rightarrow j}$ consists of an
additive, length $3$ vector component,~${\bf t}^{i \rightarrow j}$,
and a multiplicative, $3 \times 3$ matrix component,~${\bf R}^{i
\rightarrow j}$.  We will refer to~${\bf t}^{i \rightarrow j}$ as the
translational component of~${\cal T}^{i \rightarrow j}$ and to~${\bf
R}^{i \rightarrow j}$ as the rotational component of~${\cal T}^{i
\rightarrow j}$, although in general ~${\bf R}^{i \rightarrow j}$ may
be any invertible $3 \times 3$ matrix transformation.

A point, ${\bf x}^i$, expressed in the \ith coordinate system may then be
transformed to the \jth coordinate system by
\begin{equation} \label{eq:genxf}
	{\bf x}^j = 
	{\bf R}^{i \rightarrow j} {\bf x}^i + {\bf t}^{i \rightarrow j} .
\end{equation}
A transformation from the \ith coordinate system to the \jth coordinate
system may be inverted so that given~${\cal T}^{i \rightarrow j}$,
${\cal T}^{j \rightarrow i}$ may be computed by
\begin{eqnarray}
	{\bf R}^{j \rightarrow i} &=& ({\bf R}^{i \rightarrow j})^{-1} \\
	{\bf t}^{j \rightarrow i} &=& ({\bf R}^{i \rightarrow j})^{-1} (-{\bf t}^{i \rightarrow j}) ,
\end{eqnarray}
where~$(\cdot)^{-1}$~indicates~matrix~inverse.%


Because in general the bodies are in motion with respect to each other
and the world coordinate system, the transformations between
coordinate systems change over time.  We assume that the motion
data is sampled at $n$ discrete moments in time called frames, and use
${\cal T}^{i \rightarrow j}_k$ to refer to the value of ${\cal T}^{i
\rightarrow j}$ at frame $k \in [0..n-1]$.  

\figureTop{
  \centerline{\includegraphics[width= 0.75 \columnwidth]{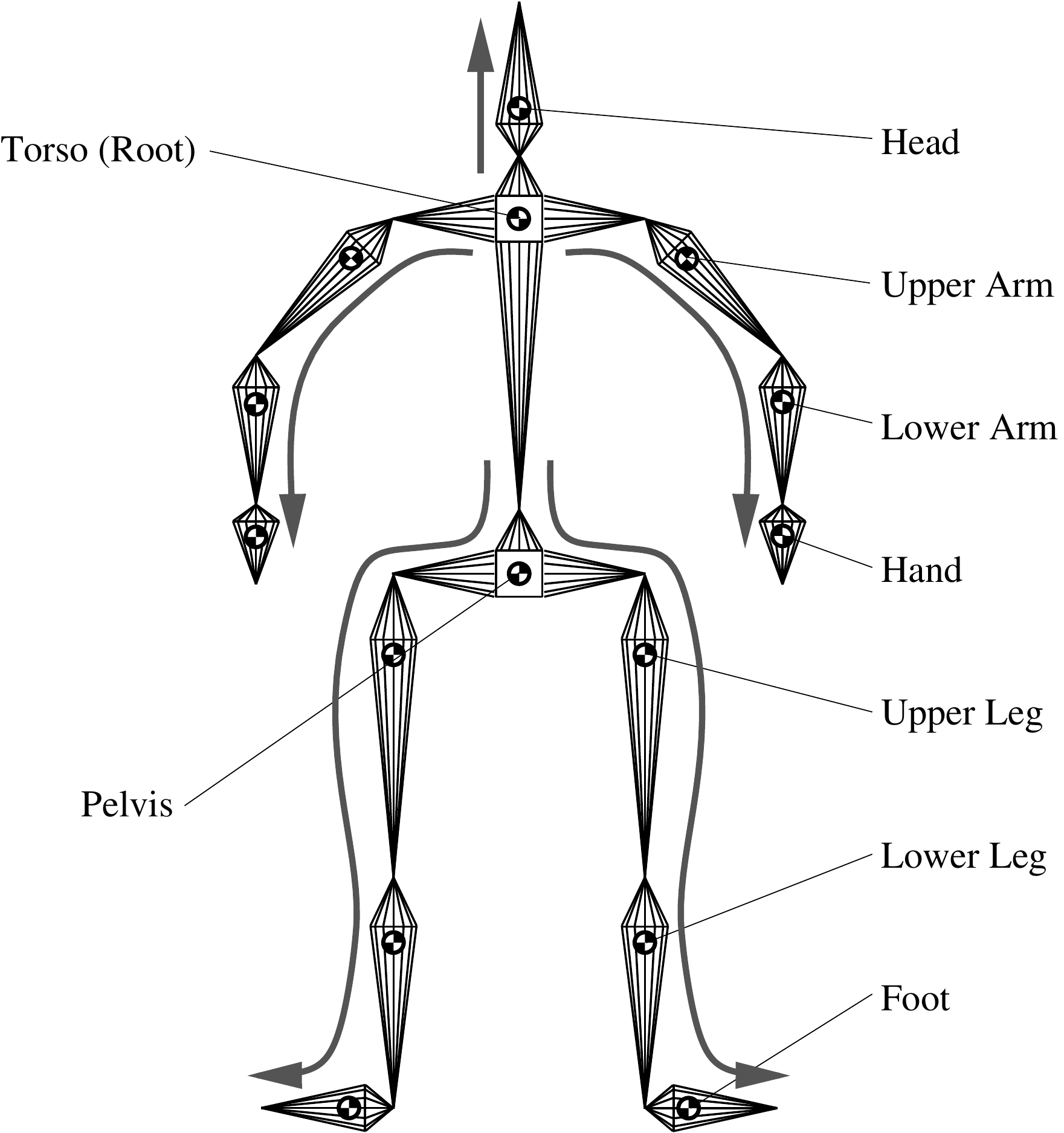}}
  \vskip -0.05in
  \Caption{
     Example of an articulated hierarchy that could be used to model
     a human figure.  The torso is the root body and the arrows
     indicate the outboard direction.  For rendering, the skeleton
     model shown here would be replaced with a more realistic graphical
     model.
  }\label{fig:ArticulatedTree}
  \vskip -0.15in
}

An articulated hierarchy is described by the topological information
indicating which bodies are connected to each other and by geometric
information indicating the locations of the connecting joints.  The
topological information takes the form of a tree%
\footnote{\mbox{We discuss the topological cycles created by loop joints in
Section~\ref{sec:Discussion}.}}  with a single body located at its
root and all other bodies appearing as nodes within the tree as shown
in Figure~\ref{fig:ArticulatedTree}.  When referring to directions
relative to the arrangement of the tree, the \textit{inboard}
direction is towards the root, and the \textit{outboard} direction is
away from the root.  Thus for a joint connecting two bodies, $i$ and
$j$, the parent body, $j$, is the inboard body and the child, $i$, is
the outboard body.  Similarly, a joint which connects a body to its
parent is that body's inboard joint and a joint connecting the body to
one of its children is an outboard joint.  All bodies have at most one
inboard joint but may have multiple outboard joints.

The hierarchy's topology is defined using a mapping function,
$P(\cdot)$, that maps each body to its parent body so that $P(i) = j$
will imply that the \jth body is the immediate parent of the
\ith body in the hierarchical tree.  The object, $\tau \in [ 0..m-1
]$, with $P(\tau) = \omega$ is the root object.  To simplify
discussion, we will temporarily assume that $P(\cdot)$ is known {\it
a~priori}.  Later, in Section~\ref{ssec:Hierarchy}, we will show how
$P(\cdot)$ may be inferred when only the ${\cal T}^{i \rightarrow
\omega}$'s are known.

\figureTop{
  \centerline{\includegraphics[width=30mm]{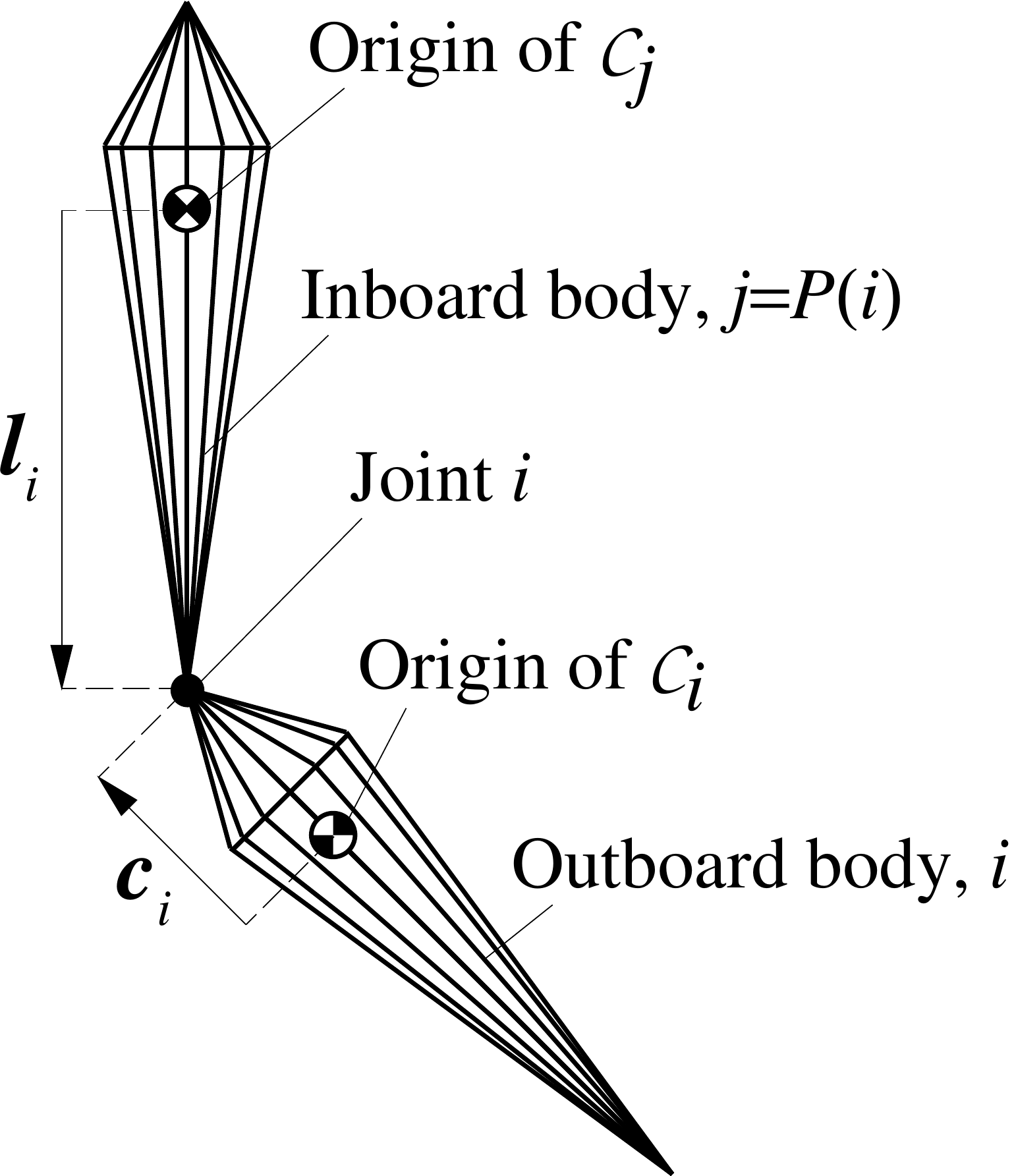}}
  \vskip -0.05in
  \Caption{
     Joint diagram showing the location of the rotary joint
     between bodies $i$ and $j=P(i)$.  The location of the joint is
     defined by a vector displacement, ${\bf c_i}$, relative to the
     coordinate system of body $i$, and a second vector displacement,
     ${\bf l_i}$, in the coordinate system of body $j$.
  }\label{fig:JointDiagram}
  \vskip -0.2in
}

The geometry of the articulated hierarchy is determined by specifying
the location of each joint in the coordinate frames of both its
inboard body and its outboard body.  Because each body has a single
inboard joint, we will index the joints so that the \ith joint is
the inboard joint of the \ith body as shown in
Figure~\ref{fig:JointDiagram}.

Let ${\bf c}_{i}$ refer to the location of the \ith joint in the
\ith body's (the joint's outboard body) coordinate system, and let
${\bf l}_{i}$ refer to the location of the \ith joint in the
$P(i)$-th body's (the inboard body's) coordinate system (see
Figure~\ref{fig:JointDiagram}).  The transformation of
equation~(\ref{eq:genxf}) that goes from the \ith coordinate system
to its parent's, $P(i)$, coordinate system can then be re-expressed in
terms of the joint locations, ${\bf c}_{i}$ and ${\bf l}_{i}$, and the
rotation at the joint, ${\bf R}^{i \rightarrow P(i)}$, so that
\begin{eqnarray}
   {\bf x}^{P(i)} &=& {\bf R}^{i \rightarrow P(i)}_k ( {\bf x}^{i} -                                {\bf c}_{i} ) + {\bf l}_{i} \\
                  &=& {\bf R}^{i \rightarrow P(i)}_k   {\bf x}^{i} - {\bf R}^{i \rightarrow P(i)}_k {\bf c}_{i}   + {\bf l}_{i} .   \label{eq:artxf}
\end{eqnarray}


\subsection{Finding Joint Locations} \label{ssec:JointLoc}

The general transformation given by equation~(\ref{eq:genxf}) applies
to any arbitrary hierarchy of bodies.  When the bodies
are connected by rotary joints, the relative motion of two connected
bodies must satisfy a constraint that prevents the joint between them
from coming apart.  Comparing equation~(\ref{eq:artxf}) with
equation~(\ref{eq:genxf}) shows that although rotational terms are the
same, the translational term of equation~(\ref{eq:genxf}) has been
replaced with the constrained term, 
$-{\bf R}^{i \rightarrow P(i)}_k{\bf c}_{i}+{\bf l}_{i}$.  
Using equation~(\ref{eq:artxf}) to transform the
location of ${\bf c}_{i}$ to the $P(i)$-th coordinate system will
identically yield ${\bf l}_{i}$, and equation~(\ref{eq:artxf})
enforces the constraint that the joint stay together.

The input transformations for each of the body parts do not
contain any explicit information about joint constraints. However, if the
motion was created by an articulated system, then it should
be possible to express the same transformations hierarchically using
equation~(\ref{eq:artxf}) and an appropriate choice of ${\bf c}_{i}$
and ${\bf l}_{i}$ for each of the joints.  Thus for each pair of
parent and child bodies, $i \neq \tau$ and $j = P(i)$, there should be
a ${\bf c}_{i}$ and ${\bf l}_{i}$ such that equation~(\ref{eq:genxf})
and equation~(\ref{eq:artxf}) are equivalent and
\begin{multline}\label{eq:tmp1}
{\bf R}^{i \rightarrow P(i)}_k   {\bf x}^{i} + {\bf t}^{i \rightarrow P(i)}_k = \\
{\bf R}^{i \rightarrow P(i)}_k   {\bf x}^{i} - {\bf R}^{i \rightarrow P(i)}_k {\bf c}_{i} + {\bf l}_{i}
\end{multline}  
for all $k \in [0..n-1]$. After simplifying, equation~(\ref{eq:tmp1}) becomes
\begin{eqnarray} \label{eq:constBasic} 
   {\bf t}^{i \rightarrow P(i)}_k = - {\bf R}^{i \rightarrow P(i)}_k {\bf c}_{i} + {\bf l}_{i} 
\end{eqnarray}  
for all $k \in [0..n-1]$.  Later, it will be more convenient to work
with the global transformations.  By applying ${\cal T}^{P(i)
\rightarrow \omega}$ to both sides of equation~(\ref{eq:constBasic})
and simplifying the result, we have
\begin{equation} \label{eq:const}
	{\bf R}^{i    \rightarrow \omega}_{k} {\bf c}_{i} + {\bf t}^{i    \rightarrow \omega}_{k} = 
	{\bf R}^{P(i) \rightarrow \omega}_{k} {\bf l}_{i} + {\bf t}^{P(i) \rightarrow \omega}_{k} 
\end{equation}
for all $k \in [0..n-1]$. Equation~(\ref{eq:const}) has a consistent
geometric interpretation: the location of the joint in
the outboard coordinate system, ${\bf c}_{i}$, and the location of the
joint in the inboard coordinate system, ${\bf l}_{i}$, should
transform to the same location in the world coordinate system; in
other words, the joint should stay together.

Equation~(\ref{eq:const}) can be rewritten in matrix form as
\begin{equation} \label{eq:constMatrix}
	{\bf Q}^{i \rightarrow P(i)}_k {\bf u}_{i} = {\bf d}^{i \rightarrow P(i)}_k .
\end{equation}
where ${\bf d}^{i \rightarrow P(i)}_k$ is the length $3$ vector given by
\begin{equation}
	{\bf d}^{i \rightarrow P(i)}_k = - ( {\bf t}^{i \rightarrow \omega}_{k} \! - {\bf t}^{P(i) \rightarrow \omega}_{k} ) ,
\end{equation}
${\bf u}_{i}$ is the length $6$ vector
\begin{equation}
	{\bf u}_{i} = \left[ 
                        \begin{array}{c}
                          {\bf c}_{i} \\
                          {\bf l}_{i}
                        \end{array}
                      \right] ,
\end{equation}
and ${\bf Q}^{i \rightarrow j}_k$ is the $3 \times 6$ matrix
\begin{equation} 
	{\bf Q}^{i \rightarrow j}_k = 
	\left[ 
	({\bf R}^{i \rightarrow \omega}_{k}) \; ({\bf -R}^{P(i) \rightarrow \omega}_{k})
	\right] .
\end{equation}

Assembling equation~(\ref{eq:constMatrix}) into a single linear system
of equations for all $0..n-1$ frames gives:

\begin{equation} \label{eq:ls-sys1}
	\left[ 
	  \begin{array}{c}
	    {\bf Q}^{i \rightarrow P(i)}_{0} \\
	    \vdots  \\
	    {\bf Q}^{i \rightarrow P(i)}_{k} \\
	    \vdots  \\
	    {\bf Q}^{i \rightarrow P(i)}_{n-1} 
	  \end{array}
	\right] \, 
        \left[ 
	  \begin{array}{c}
            {\bf c}_{i} \\
            {\bf l}_{i} 
	  \end{array}
	\right] =
	\left[ 
	  \begin{array}{c}
	    {\bf d}^{i \rightarrow P(i)}_{0} \\
	    \vdots  \\
	    {\bf d}^{i \rightarrow P(i)}_{k} \\
	    \vdots  \\
	    {\bf d}^{i \rightarrow P(i)}_{n-1} 
	  \end{array}
	\right] .
\end{equation}	
The matrix of ${\bf Q}$'s is $3n \times 6$ and will be denoted by
$\widehat{\bf Q}$; the matrix of ${\bf d}$'s is $3n \times 1$. The
linear system of equations in equation~(\ref{eq:ls-sys1}) can be used
to solve for the joint location parameters, ${\bf c}_{i}$ and ${\bf
l}_{i}$.

Unless the input motion data consists of only two frames of motion,
$\widehat{\bf Q}$ will have more rows than columns and the system
will, in general, be over-constrained.  Nonetheless, if the motion was
generated by an articulated model, an exact solution will exist.
Realistically, limited sensor precision and other sources of error
will prevent an exact solution, and a best-fit solution must be found
instead.

Despite the fact that the system will be over-constrained, it may be
simultaneously under-constrained if the input motions do not span the
space of rotations.  In particular, if two bodies connected by a joint
do not rotate with respect to each other, or if they do so but only
about a single axis, then there will be no unique answer.  In the case
where they are motionless with respect to each other, any location
in space would be a solution.  Similarly, if their relative rotations
are about a single axis, then any point on that axis could serve as
the joint's location.  For reasons of numerical accuracy, in either of
these cases the desired solution is chosen to be the one closest to
the origin of the inboard and outboard body coordinate frames.

The technique of solving for a least squares solution using the
singular value decomposition is well suited for this type of
problem\cite{Press:1994:nrc}.  Because there is no numerical structure
in our problem that we can exploit (such as sparsity), our use of this
technique to solve equation~(\ref{eq:ls-sys1}) is straightforward. In
later sections, we will use the residual vector from the solution of
this system to show the translational difference between the input
data and the value given by equation~(\ref{eq:artxf}).

\subsection{Single-axis Joints} \label{ssec:JointAxis}

If a joint rotates about two or more non-parallel axes, enough
information is available to resolve the location of the joint center
as described above.  However, if the joint
rotates about a single axis, then a unique joint center does not
exist, and any point along the axis is an equally good solution to
equation~(\ref{eq:ls-sys1}).  In these cases the solution to
equation~(\ref{eq:ls-sys1}) found by the singular value decomposition
will be an essentially arbitrary point on the axis.

This situation can be detected by examining the singular values of
$\widehat{\bf Q}$ from equation~(\ref{eq:ls-sys1}).  If one of the
singular values of $\widehat{\bf Q}$ is near zero, i.e., if
$\widehat{\bf Q}$ is rank deficient, then that joint is a single-axis
joint, or at least in the input motion it rotates only about a single
axis.  The first three components of the corresponding column vector
of ${\bf V}$ from the singular value decomposition are the joint axis
in the inboard coordinate frame; the second three are the axis in the
outboard coordinate frame

While we were able to verify this method for detecting single-axis
joints using synthetic data, none of the data from our motion capture
trials indicated the presence of a single-axis joint.  We believe that
single-axis joints did not appear in the data because our subjects were 
specifically asked to exercise the full range of motion and all degrees of 
freedom of their joints.  As a result, the system was able to determine a 
location even for joints such as the knee and elbow that are traditionally 
approximated as one degree-of-freedom joints.


\subsection{Determining the Body Hierarchy} \label{ssec:Hierarchy}

In the previous sections, we assumed that the hierarchical
relationship between the bodies given by the parent function,
$P(\cdot)$, is known.  In some instances, however, determining a
suitable hierarchy automatically by inferring it from the input
transformation matrices may be desirable.  Our algorithm does this by
finding the parent function that minimizes the sum of the
$\varepsilon_i$'s for all the joints in the hierarchy.

The problem of finding the optimal hierarchy is equivalent to finding
a minimal spanning tree.  Each body can be thought of as a node,
joints are the edges between them, and the joint fit error,
$\varepsilon_i$, is the weight of the edge.  The hierarchy can then be
determined by evaluating the joint error between all pairs of bodies,
selecting a root node, $\tau$, and then constructing the minimal
spanning tree.  See~\cite{Cormen:1991:ita} for example algorithms.


\subsection{Removing the Residual} \label{ssec:Reconstruction}

After we have determined the locations of the joints, we can use this
information to construct a model that approximates the dimensions of
the subject.  This model can then be used to play back the original
motion data.  Unless the residual errors on the joint fits were all
near zero, the motion will cause the joints of the model to move apart
from each other during playback in a fashion that is typical of
unconstrained motion capture data.  If, however, we use the inferred
joint locations to create an articulated model with kinematic joint
constraints and then play back the motion through this model, the
joints will stay together.  Playing back motion capture data by
applying only the rotations to an articulated model is common
practice; the difference here is that the model itself has been
generated from the motion data.  Essentially, we have projected the
motion data onto a parametric model and then used the fit to discard
the residual.

 
\section{Results}

To verify that our algorithm could be used to determine the hierarchy
and joint parameters from motion data, we tested it on both simulated
data and on data acquired from a magnetic motion capture system.
First, the technique was tested on a rigid-body dynamic simulation of a human
containing 48 degrees of freedom.  The simulated figure was moved so
that all of its degrees of freedom were exercised. The algorithm
correctly computed limb lengths within the limits of numerical
precision (errors less than $\text{10}^{\text{-6}}$\,m) and determined
the correct hierarchy.

We next tested our method in a magnetic motion capture environment.
Magnetic motion capture systems are frequently noisy, and the
Ascension system we used has a published resolution of about
4\,mm\cite{Ascension}.  To establish a baseline for the amount of
noise present in the environment, two sensors were rigidly attached
56.5\,cm~apart and moved through the capture space. The results of
this experiment are shown in Figure~\ref{fig:Calibration}.  A scale
factor exists when converting from units the motion capture system
reports to centimeters, and we calculated this scale factor to be 0.94
based on the mean of this data set.  The scaled standard deviation of
the data is 0.7\,cm.

\figureTop{
  \centerline{\includegraphics[width= .9 \columnwidth]{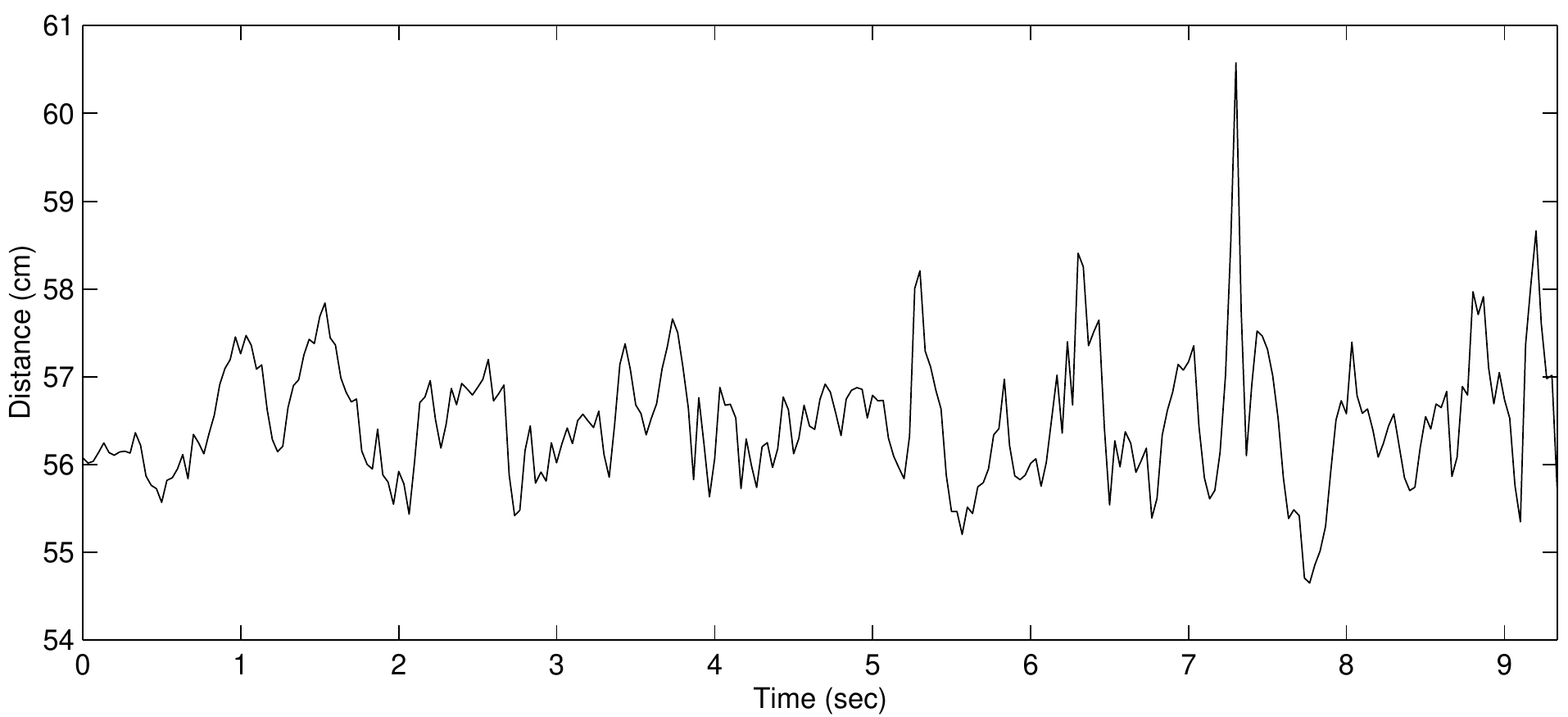}}
  \vskip -0.1in
  \Caption{ 
    Calibration data showing the distance between two markers
    attached rigidly to one another and moved through the capture
    space. If the sensors are not moved, the data is much less
    noisy.  
  }\label{fig:Calibration}
  \vskip -0.2in
}

To test the algorithm on something less complicated than biological
joints, we constructed a wooden mechanical linkage with five
ball-and-socket joints. That linkage is shown in
Figure~\ref{fig:FredPicture}. Six sets of data were captured in which
all the degrees of freedom were exercised.  Before Set~6 was captured,
the marker positions were moved to evaluate the robustness of the
method to changes in marker locations.  The results are shown in
Table~\ref{table:FredTable} along with the measured values of the
joint-to-joint distances. The maximum error across all trials is
1.1\,cm, and the hierarchy was computed correctly for each trial.
Another way of evaluating the fit is to examine the residual vectors
from the least squares process.  The norms of the residual vectors for
the best fit (Set 1, Right Shoulder) and the worst fit (Set 6, Left
Shoulder) are shown in Figures~\ref{fig:residf1}
and~\ref{fig:residf2}, respectively.  The right-hand graph has an
asymmetric distribution because it is the distribution of an absolute
value. We regard these results as very good because the error is on
the order of the resolution of the sensors.

\figureTopBot{
  \centerline{\includegraphics[width=\columnwidth]{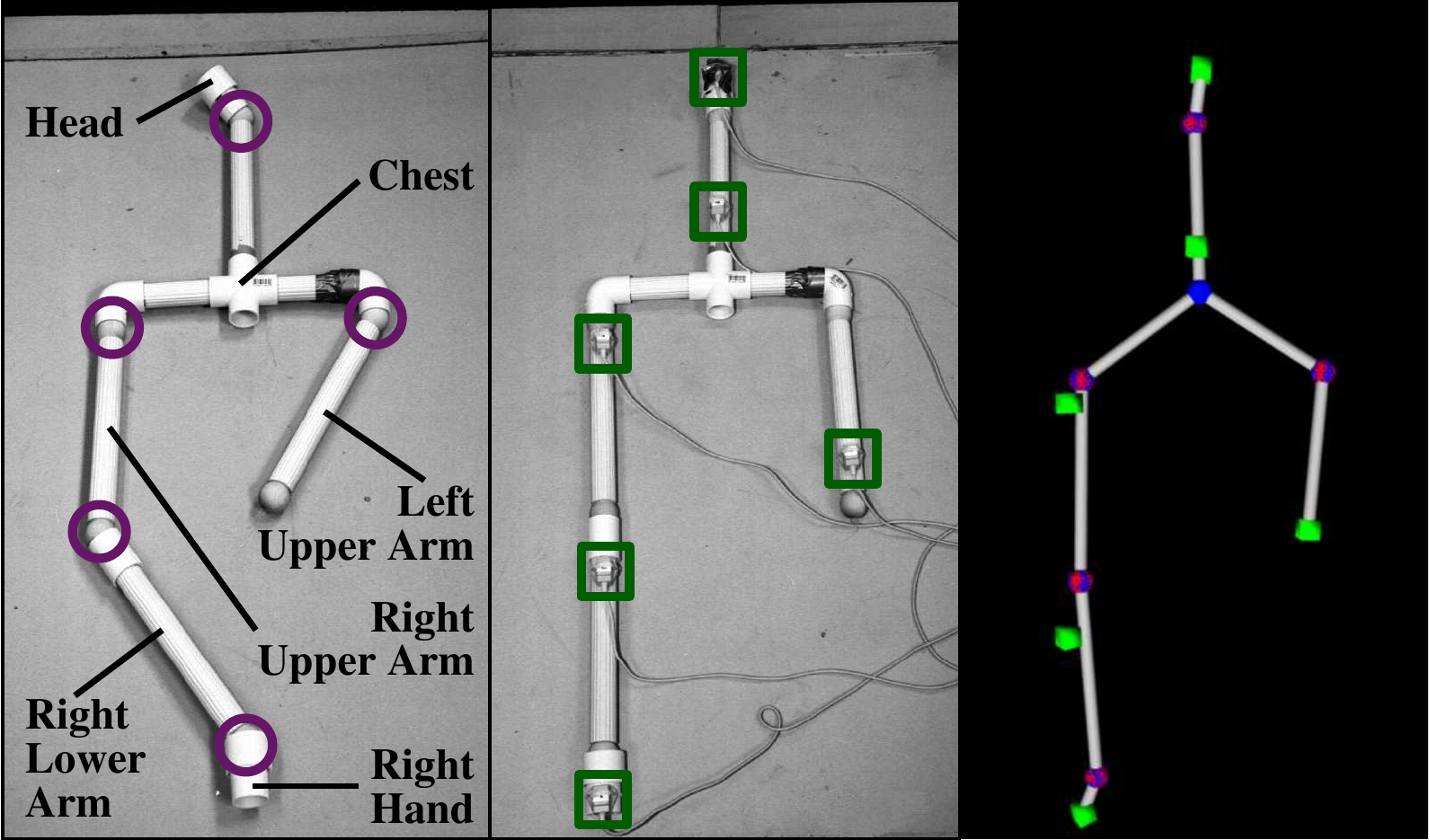}}
  \vskip -0.05in
  \centerline{\hfill(A)\hfill\hfill(B)\hfill\hfill(C)\hfill}
  \vskip -0.05in
  \Caption{
    Wooden mechanical linkage. (A)~Labels indicate the terms
    that we used to refer to the body parts; circles
    highlight the joint locations.  (B)~The motion capture sensors
    (highlighted squares) have been attached to the linkage.
    (C)~The model computed automatically from the motion data using
    our algorithm.  The joints are shown with  spheres, and the
    sensors with  cubes.  Links between joints are indicated with
    cylinders.
  }\label{fig:FredPicture}
  \vskip -0.1in
}

\begin{table*}
%
%
  \begin{center}\footnotesize
    \begin{tabular}{||l||c||c|c|c|c|c|c||c|c|c|c|c|c||}
      \hline
      \rule{0ex}{2ex}           & Meas. & Set 1 & Set 2 & Set 3 & Set 4 & Set 5 & Set 6 & $\Delta$~1 & $\Delta$~2 & $\Delta$~3 & $\Delta$~4 & $\Delta$~5 & $\Delta$~6 \\
      \hline
      \hline
      Neck --- Left Shoulder    & 39.0 & 39.4 & 38.8 & 39.8 & 39.1 & 39.1 & 40.1 & -0.4 & \hphantom{-}0.2 & -0.8 & -0.1 & -0.1 & -1.1 \\
      \hline 								 
      Neck --- Right Shoulder   & 39.7 & 39.8 & 39.8 & 40.3 & 40.0 & 39.9 & 40.3  & -0.1 & -0.1 & -0.6 & -0.3 & -0.2 & -0.6 \\
      \hline								 
      Between Shoulders         & 34.3 & 34.3 & 33.7 & 34.5 & 34.3 & 34.3 & 34.8  & \hphantom{-}0.0 & \hphantom{-}0.6 & -0.2 & \hphantom{-}0.0 & \hphantom{-}0.0 &  -0.5    \\
      \hline								 
      Right Upper Arm           & 28.6 & 29.2 & 29.0 & 28.8 & 28.9 & 29.0 & 29.1  & -0.6 & -0.4 & -0.2 & -0.3 & -0.4 & -0.5 \\
      \hline								 
      Left Upper Arm            & 31.4 & 31.5 & 31.7 & 31.9 & 31.5 & 31.1 & 31.2  & -0.1 & -0.3 & -0.5 & -0.1 & \hphantom{-}0.3 & \hphantom{-}0.2   \\
      \hline
     \end{tabular}
  \end{center}
  \vskip -0.15in
  \Caption{
    A comparison of measurements and calculated limb lengths
    for six data sets of the mechanical linkage. The units are centimeters and
    the columns labeled $\Delta$ show the difference in measured and
    calculated values.  Joint names follow the analogy with human
    physiology used in Figure~\protect{\ref{fig:FredPicture}}(A). 
  }\label{table:FredTable}
  \vskip -0.2in
\end{table*}

\figureTopBot{
  \centerline{\includegraphics[width=\linewidth]{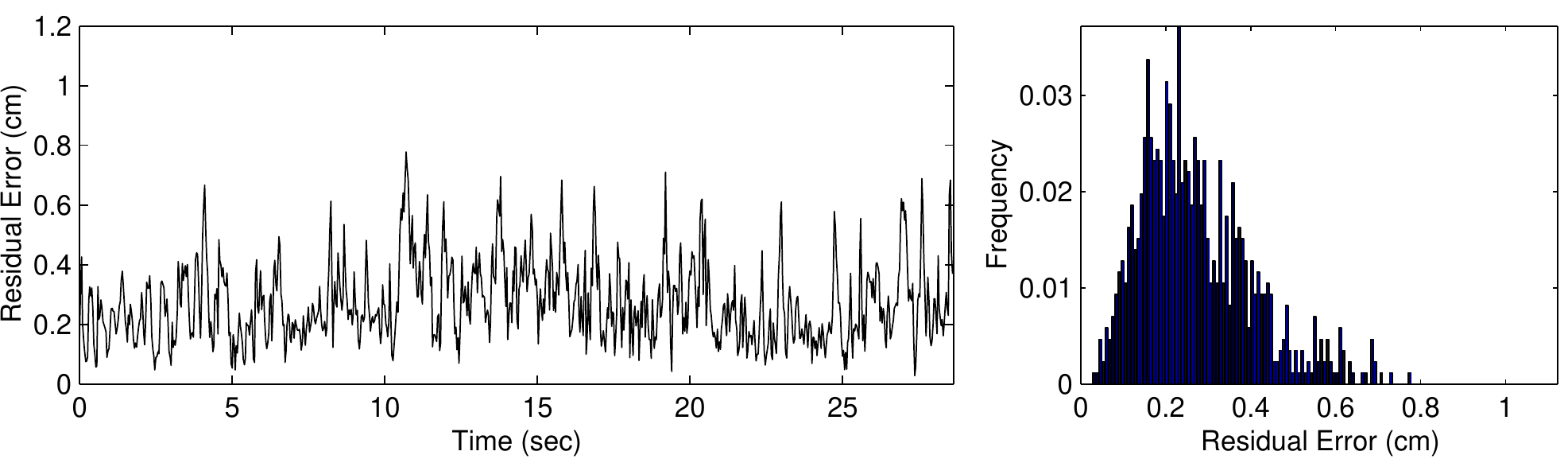}}
  \vskip -0.05in
  \Caption{
    Residual errors of the right shoulder joint for the data
    from Set~1 for the mechanical linkage
    (Table~\protect{\ref{table:FredTable}}). The left graph shows the
    magnitude of the residual vector. The right graph shows the
    distribution of the frequency of the magnitudes.
  }\label{fig:residf1}
  \vskip -0.1in
}

\figureTopBot{
  \centerline{\includegraphics[width=\linewidth]{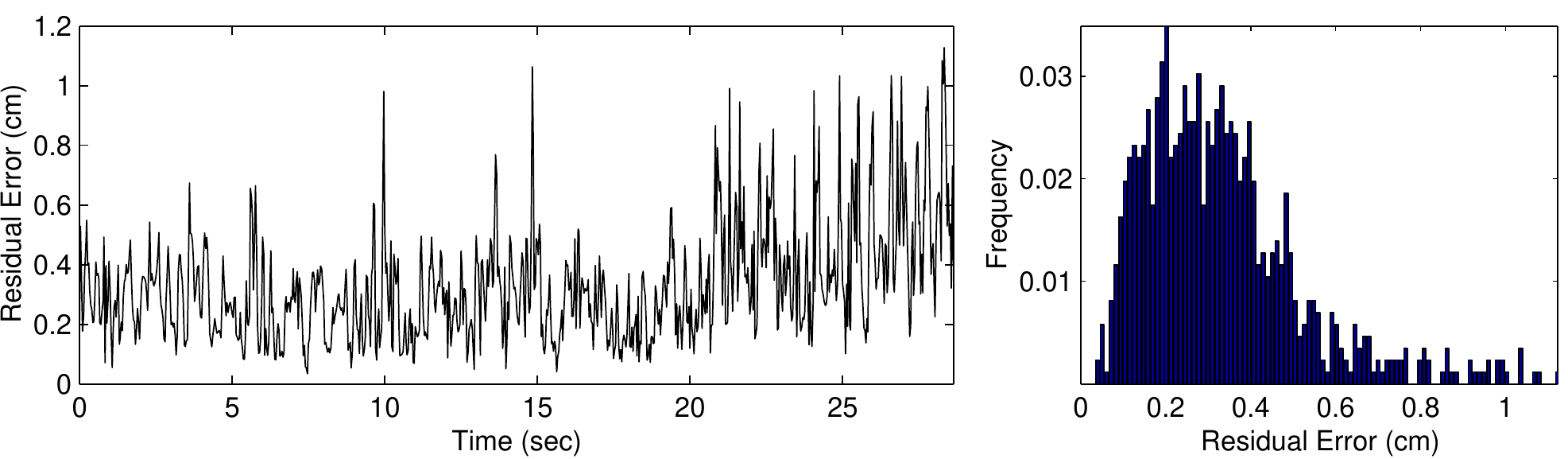}}
  \vskip -0.05in
  \Caption{
    Residual errors of the left shoulder joint for the data
    from Set~6 for the mechanical linkage
    (Table~\protect{\ref{table:FredTable}}).  
  }\label{fig:residf2}
  \vskip -0.2in
}

The important test case, of course, is to verify that we are able to
estimate the limb lengths of people. This task is more difficult
because human joints are not simple mechanical linkages. To provide a
basis for comparison, we measured the limb lengths of our test
subjects. As mentioned previously, this process is inexact and prone
to error, but it does provide a plausible estimate.  We measured limb
lengths from bony landmark to bony landmark to provide repeatability
and consistency in our measurements. For example, the upper leg of a
subject was measured as the distance from the top of the greater
trochanter of the femur to the lateral condyle of the tibia. Because
the head of the femur extends upward and inward into the innominate,
this measurement will be inaccurate by a few centimeters.
Nonetheless, because the greater trochanter is the only palpable area
at the upper end of the femur, this measurement is the best available.
The difficulty in obtaining accurate hand measurements is one of the
primary reasons that we chose to develop our automatic technique.

Our test subjects performed two different sets of motions for capture.
We refer to the first set as the ``exercise'' set. In it the subjects
attempted to move each joint essentially in isolation to generate a
full range of motion for each joint.  Thus the routine consists of a
set of discrete motions such as rolling the head around on the neck,
bending at the waist, high-stepping, lifting each leg and waving it
about, lifting the arms and waving them about, bending the elbows and
the wrists, etc. This exercise set mimics the way we gathered data for
the mechanical linkage. We refer to the second set of motions captured
as the ``walk'' sets. In it the subjects try to move as many degrees
of freedom at once as they can while walking. This routine is perhaps
best described as a ``chicken'' walk, consisting of highly exaggerated
leg movements coupled with bending the waist and waving the arms
about.

A male subject performed the two types of motion and the results of
the limb length calculations are shown in Tables~\ref{table:gexer}
and~\ref{table:gwalk}.  As expected, the residual errors for a human
are much larger than for the mechanical linkage. A representative
example is shown in Figure~\ref{fig:residg}. For this subject, the
maximum difference between measured and calculated values is 4.1\,cm,
and occurs at the left upper arm during one of the exercise sets.  The
mean of the differences between calculated and measured values is less
than one centimeter for every limb except the upper arms where it is
1.4\,cm and 2.2\,cm for the right and left arms, respectively.  The
algorithm consistently finds a longer length for the left upper arm
than what we measured, and that difference may be due, in part, to an
error in the value measured by hand.  However, the shoulder joint is
poorly approximated by a rotary joint: an accurate biomechanical
rigid-body model would have at least seven degrees of
freedom\cite{Vanderhelm:1994:fem, Vanderhelm:1994:akd}, and it is not
surprising that the worst fit occurs there.

The same motions were repeated with a female subject, and the
results are shown in Table~\ref{table:cnd}.  The largest difference
between calculated and measured values is 2.4\,cm and again occurs
for the left upper arm.  The algorithm also finds a longer length
for the left upper arm than we measured.  The maximum error is less
than that for the male test subject, but less consistency was found
among the results for the female test subject.  The mean of the
differences between the calculated and measured values is greater than
one centimeter for the right lower leg, left upper leg, and left upper
arm.

The system also computed a hierarchy for each trial.  For all
``exercise'' trials for both male and female subjects the computed
hierarchy was correct; however, results from the ``walk'' data were
less satisfactory.  For three of the five ``walk'' trials, the
algorithm improperly made one of the upper legs a child of the other
instead of the pelvis.  This error may have occurred because the
pelvis sensor was mounted on the system's battery pack worn on the
subject's hip.  Motion in this sensor caused by rotating the thigh upwards 
may have contributed to the error.  The limb length
results we report are, of course, for the correct hierarchy
assignments.

In addition to the joint measurements we reported, our algorithm
determines information for joints (such as between the chest and
pelvis) that model the bending of the torso but which are gross
approximations to the way the human spine bends.  Our algorithm
reports limb lengths for these joints within the torso, and these are
generally consistent with the dimensions of the torsos of the
subjects.  However, because we have no reasonable way of measuring
these lengths for comparison, we have omitted them from the results.
The locations computed for these joints can be seen in
Figure~\ref{fig:PersonFigure} and in the animations that accompany
this paper.
 
The algorithm is quite fast. On an SGI O2 with a 195 MHz R10000
processor, less than 4~seconds are required to process 45~seconds of
motion data for 16~sensors with the hierarchy specified, and less than
14~seconds when the hierarchy was not specified.

\begin{table*}[!t]
  \begin{center}\footnotesize
    \begin{tabular}{||l||c||c|c|c|c||c|c|c|c||}
      \hline
      \rule{0ex}{2ex} & Meas. & Exer. 1 & Exer. 2 & Exer. 3 & Exer. 4 & $\Delta$ 1 & $\Delta$ 2 & $\Delta$ 3 & $\Delta$ 4\\
      \hline
      \hline
      Right Lower Leg & 40.0 & 40.8 & 40.9 & 42.2 & 42.5 & -0.8 & -0.9 & -2.2 & -2.5 \\  
      \hline                                                                           
      Left Lower Leg  & 40.3 & 37.3 & 38.4 & 41.2 & 41.5 & \hphantom{-}3.0 & \hphantom{-}1.9 & -0.9 & -1.2 \\    
      \hline                                                                           
      Right Upper Leg & 41.6 & 41.5 & 42.1 & 42.9 & 42.2 & \hphantom{-}0.1 & -0.5 & -1.3 & -0.6 \\ 
      \hline                                                                           
      Left Upper Leg  & 43.2 & 41.4 & 41.8 & 43.2 & 43.0 & \hphantom{-}1.8 & \hphantom{-}1.4 & \hphantom{-}0.0 & \hphantom{-}0.2 \\        
      \hline                                                                           
      Right Lower Arm & 27.0 & 26.3 & 26.7 & 27.7 & 27.0 & \hphantom{-}0.7 & \hphantom{-}0.3 & -0.7 & \hphantom{-}0.0 \\         
      \hline                                                                           
      Left Lower Arm  & 26.7 & 26.5 & 27.0 & 26.7 & 27.1 & \hphantom{-}0.1 & -0.3 & -0.1 & -0.4 \\   
      \hline                                                                           
      Right Upper Arm & 29.5 & 32.1 & 31.3 & 29.3 & 28.8 & -2.6 & -1.8 & \hphantom{-}0.2 & \hphantom{-}0.7 \\  
      \hline                                                                           
      Left Upper Arm  & 29.5 & 33.7 & 32.9 & 30.1 & 29.9 & -4.1 & -3.4 & -0.6 & -0.4 \\
      \hline                                               
    \end{tabular}
  \end{center}
  \vskip -.15in
  \Caption{
    A comparison of measurements and calculated limb lengths
    for four data sets of a male  subject attempting to exercise
    each degree of freedom essentially in isolation. 
  }\label{table:gexer}

  \begin{center}\footnotesize
    \begin{tabular}{||l||c||c|c|c||c|c|c||}
      \hline
      \rule{0ex}{2ex} & Meas. & Walk 1 & Walk 2 & Walk 3 & $\Delta$ 1 & $\Delta$ 2 & $\Delta$ 3\\
      \hline
      \hline
      Right Lower Leg & 40.0 & 40.7 & 40.3 & 38.9 & -0.6 & -0.3 & \hphantom{-}1.1 \\   
      \hline                                                             
      Left Lower Leg  & 40.3 & 40.8 & 38.9 & 39.8 & -0.4 & \hphantom{-}1.4 & \hphantom{-}0.5 \\  
      \hline                                                             
      Right Upper Leg & 41.6 & 40.7 & 40.6 & 42.6 & \hphantom{-}0.9 & \hphantom{-}1.0 & -1.0 \\      
      \hline                                                             
      Left Upper Leg  & 43.2 & 45.1 & 42.7 & 43.1 & -1.9 & \hphantom{-}0.5 & \hphantom{-}0.1 \\  
      \hline                                                             
      Right Lower Arm & 27.0 & 27.3 & 27.5 & 25.8 & -0.3 & -0.5 & \hphantom{-}1.2 \\   
      \hline                                                             
      Left Lower Arm  & 26.7 & 26.2 & 24.9 & 25.6 & \hphantom{-}0.5 & \hphantom{-}1.7 & \hphantom{-}1.0 \\     
      \hline                                                             
      Right Upper Arm & 29.5 & 31 & 31.1 & 32.7 & -1.4 & -1.6 & -3.2 \\  
      \hline                                                             
      Left Upper Arm  & 29.5 & 32.3 & 32.3 & 30.8 & -2.7 & -2.7 & -1.3 \\
      \hline                                                             
    \end{tabular}
  \end{center}
  \vskip -.15in
  \Caption{
    A comparison of measurements and calculated limb lengths
    for three data sets of a male subject attempting to exercise
    all degrees of freedom simultaneously. 
  }\label{table:gwalk}

  \begin{center}\footnotesize
    \begin{tabular}{|l|c||c|c|c||c|c|c||}
      \hline
      \rule{0ex}{2ex} & Meas. & Exer. 1 & Walk 1 & Walk 2 & $\Delta_e$ 1 & $\Delta_w$ 1 & $\Delta_w$ 2\\
      \hline
      \hline
      Right Lower Leg & 36.8 & 39.1 & 38.0 & 38.1 & -2.3 & -1.2 & -1.3 \\ 
      \hline                                                            
      Left Lower Leg  & 36.5 & 37.6 & 37.0 & 37.4 & -1.1 & -0.5 & -0.9 \\ 
      \hline                                                            
      Right Upper Leg & 42.2 & 42.9 & 43.3 & 42.2 & -0.7 & -1.1 & \hphantom{-}0.0 \\  
      \hline                                                            
      Left Upper Leg  & 41.9 & 42.4 & 44.1 & 42.9 & -0.5 & -2.2 & -1.0 \\ 
      \hline                                                            
      Right Lower Arm & 24.8 & 25.5 & 25.3 & 22.4 & -0.7 & -0.5 & \hphantom{-}2.3 \\
      \hline                                                            
      Left Lower Arm  & 24.8 & 25.1 & 24.8 & 23.0 & -0.3 & \hphantom{-}0.0 & \hphantom{-}1.8 \\     
      \hline                                                            
      Right Upper Arm & 27.6 & 27.5 & 27.5 & 28.7 & \hphantom{-}0.2 & \hphantom{-}0.1 & -1.0 \\   
      \hline                                                            
      Left Upper Arm  & 27.6 & 28.5 & 30.0 & 29.0 & -0.9 & -2.4 & -1.3 \\   
      \hline                
    \end{tabular}
  \end{center}
  \vskip -.15in
  \Caption{
    A comparison of measurements and calculated limb lengths
    for four data sets of a female subject. The column labeled
    ``Exercise'' denotes a performance attempting to exercise each
    degree of freedom in isolation. Columns labeled ``Walk''
    denote a performance attempting to exercise all degrees of freedom
    simultaneously. The units are centimeters, and the columns labeled $\Delta$
    show the difference in measured and calculated values for the
    appropriate set.
  }\label{table:cnd}
  \vskip -0.2in
\end{table*}

\figureTopBot{
  \centerline{\includegraphics[width=\linewidth]{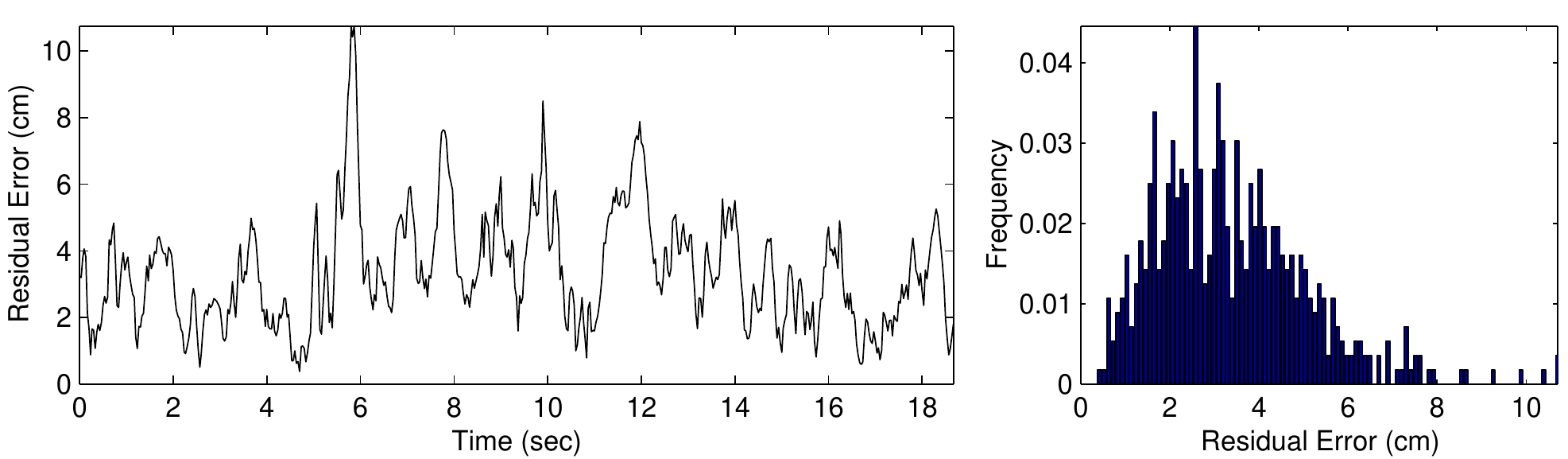}}
  \vskip -0.05in
  \Caption{
    Residual errors of the left shoulder for the data from
    Walk~2 of a male subject
    (Table~\protect{\ref{table:gwalk}}).  The
    scale of the residual vectors is larger than that of the residual
    vectors for Figures~\protect{\ref{fig:residf1}}
    and~\protect{\ref{fig:residf2}}.
  }\label{fig:residg}
  \vskip -0.2in
}

\section{Discussion and Conclusions} \label{sec:Discussion}

This paper presents an automatic method for computing limb lengths,
joint locations, and sensor placement from magnetic motion capture
data.  The method produces results accurate to the resolution of the
sensors for data that was recorded from a mechanical device
constructed with rotary joints.  The accuracy of the results for data
recorded from a human subject is consistent with estimates in the
biomechanics literature for the error introduced by approximating
human joints as rotational and assuming that the skin does not move
with respect to the bone.

Measuring and calibrating a performer in a production animation
environment is tedious.  Because this algorithm runs very quickly, it
provides a rapid way to accomplish the calibration for magnetic motion
capture systems.  Detecting and correcting for marker slippage are
additional complications in the motion capture pipeline.  Because this
technique looks for large changes in the joint residual, it provides a
rapid way of determining if a marker slipped during a particular
recorded segment, thus allowing the segment to be performed again
while the subject is still suited with sensors.

The parameters computed by this method can be used to create a digital
version of a particular performer by matching a graphical model to the
proportions of the motion capture subject.  The process does not
require the subject to assume a particular pose or to perform specific
actions other than to exercise their joints fully.  Therefore, the
method can be incorporated into applications where explicit
calibration is infeasible.  A cleverly disguised ``exercise'' routine,
for example, could be part of the pre-show portion of a location-based
entertainment experience.

The algorithm would also be of use in applications where the problem
is fitting data to a graphical model with dimensions different from
those of the motion capture subject.  The algorithm presented here
could be used in a pre-processing step to provide the best-fit limb
lengths for the data and modify the data to have constant limb
lengths.  Then constraint-based techniques could be applied to adapt
the resulting motion to the new dimensions of the graphical character.


Passive optical systems often have problems with marker identification
because occlusion causes markers to appear to swap.  For example, when
the hand passes in front of the hip during walking, the marker on the
hand and the one on the hip may become confused.  If this happens, the
marker locations may change relatively smoothly but the joint center
of the inboard and outboard bodies for each marker will change
discontinuously. This error should be identifiable when the data is
processed, allowing the markers to be disambiguated.

For relatively clean data, this algorithm can be used to extract the
hierarchy automatically.  Specifying the hierarchy is not burdensome
for magnetic motion capture data because the markers are uniquely
identified by the system. However, automatic identification of the
hierarchy might be useful in situations where connections between
objects are dynamic such as pairs dancing or a subject manipulating an
instrumented object.

We have assumed that the hierarchy is a strict tree and does not
contain cycles or loop joints such as the closed chain that is
created when the hands are clasped.  If the hierarchy is
known {\em a priori}, the location of a loop joint is found just
as it is for any other joint.  If the hierarchy is not known, the
method of Section~\ref{ssec:Hierarchy} will not find cycles and the
hierarchy it returns will be missing the additional joints required to
close the loops.  This problem could be detected by informing the user
that a joint fit with a low error was not used in building the tree.

The algorithm we have described is statistically equivalent to fitting
a parameterized model to a distribution.  The rotary joint model that
is commonly used for skeletal animation is linear, but more complex
models that explicitly model the errors introduced by the non-rotational
nature of the joints, the slippage of skin, or the noise distribution
seen in the magnetic setup would be non-linear.  Non-linear models
have been used in robotics research to model elastic deformation of
robot limb segments, joints that do not have a fixed center of
rotation, and dynamic variation due to system inertial
properties\cite{Lai:1988:idp,Goldenberg:1992:iip,Williams:1992:ika,Gourdeau:1996:pis,Karan:1994:cam}.
Reconstructing the motion based on the joint locations, as described
in Section~\ref{ssec:Reconstruction}, is a first step towards
identifying the components of the motion that are due to actual motion
and those that are due to errors.  The addition of more sophisticated
models could allow us to separate components of the data attributable
to the motion of the subject from components that are due to other
sources. This separation might allow accurate data to emerge even
from systems where the sensors are only loosely attached to the
subject.


\acknowledgements {
  \ifthenelse{\AnonymousAuthors = 0}{ 
    The authors would like to thank Victor Zordan for helping with the
    motion capture equipment and the use of his software.  Christina
    De Juan also helped with various phases of the motion capture
    process.  We also thank Len Norton for his assistance during the early
    stages of this project.

    This project was supported in part by NSF NYI Grant
    NSF-9457621, Mitsubishi Electric Research Laboratory, and a
    Packard Fellowship.  The first author was supported by a
    Fellowship from the Intel Foundation, the second author by an
    NSF CISE Postdoctoral award, NSF-9805694, and the third author
	by NSF Graduate Research Tranineeships, NSF-9454185. The motion 
	capture system was purchased with funds from NSF Instrumentation Grant,
    NSF-9818287.
  }{
    Deleted for anonymous review.
  }
}


\bibliography{motmot}


\end{document}